\documentclass[acmsmall]{acmart}

\AtBeginDocument{%
	\providecommand\BibTeX{{%
			\normalfont B\kern-0.5em{\scshape i\kern-0.25em b}\kern-0.8em\TeX}}}

\usepackage{graphicx}
\usepackage{graphics}
\usepackage{graphics,color}
\usepackage{url}
\usepackage{epstopdf}
\usepackage[T1]{fontenc}
\usepackage[utf8]{inputenc}
\usepackage{times}
\usepackage{times}
\usepackage{tabularx}
\usepackage{tikz}
\usetikzlibrary{matrix}
\usepackage{cite}
\usepackage{ifthen}
\usepackage{times}
\usepackage{tabularx}
\usepackage{epstopdf}
\usepackage{bm}

\newcommand{\hide}[1]{\ifthenelse{\boolean{false}}{#1}{}}

\newtheorem{theorem}{{\bf Theorem}}
\newtheorem{lemma}{{\bf Lemma}}
\newtheorem{claim}{{\bf Claim}}

\newtheorem{corollary}{{\bf Corollary}}

\newtheorem{remark}{{\bf Remark}}

\newtheorem{defn}{Definition}

\newtheorem{ex}{Example}

\newcommand{\barr}{\begin{array}}
	\newcommand{\earr}{\end{array}}

\newcommand{\benum}{\begin{enumerate}}
	\newcommand{\eenum}{\end{enumerate}}

\newcommand{\bit}{\begin{itemize}}
	\newcommand{\eit}{\end{itemize}}

\newcommand{\bdes}{\begin{description}}
	\newcommand{\edes}{\end{description}}

\newcommand{\bfig}{\begin{figure}}
	\newcommand{\efig}{\end{figure}}

\newcommand{\bemq}{\begin{quote} \begin{em}}
		\newcommand{\eemq}{\end{em} \end{quote}}

\newcommand{\cbrac}[1]{\left\{{#1}\right\}}

\newcommand{\indic}[1]{I_{\cbrac{#1}}}

\newcommand{\given}{\arrowvert}

\newcommand{\ie}{{i.e.}}

\newcommand{\expect}[1]{\mathbb{E}\left[{#1}\right]}

\newcommand{\bt}{\begin{theorem}}
	\newcommand{\bl}{\begin{lemma}}
		\newcommand{\bc}{\begin{claim}}
			\newcommand{\bp}{\begin{Proposition}}
				\newcommand{\bcoro}{\begin{corollary}}
					\newcommand{\bres}{\begin{Result}}
						\newcommand{\brem}{\begin{Remark}}
							\newcommand{\et}{\end{theorem}}
						\newcommand{\el}{\end{lemma}}
					\newcommand{\ec}{\end{claim}}
				\newcommand{\ep}{\end{Proposition}}
			\newcommand{\ecoro}{\end{corollary}}
		\newcommand{\eres}{\end{Result}}
	\newcommand{\erem}{\end{Remark}}

\newcommand{\beq}{\begin{equation}}
	\newcommand{\eeq}{\end{equation}}
\newcommand{\mT}{\mathcal{T}}
\newcommand{\norm}[1]{\|{#1}\|}

\newcommand{\mb}[1]{\mathbb{#1}}
\newcommand{\mf}[1]{\mathbf{#1}}
\newcommand{\mc}[1]{\mathcal{#1}}

\begin{document}
	
	\title{Stability Analysis of a Quantum Network with Max-Weight Scheduling}
	%
		\author{Thirupathaiah~Vasantam}
		\affiliation{
			\institution{College of Information and Computer Science, University of Massachusetts}
			\city{Amherst}
			\country{USA}}
		\email{tvasantam@umass.edu}
	\author{Don Towsley}
		\affiliation{
			\institution{College of Information and Computer Science, University of Massachusetts}
			\city{Amherst}
			\country{USA}
		}
	\email{towsley@cs.umass.edu}
	%
	%
	%
	%
	%
	%
	%
	\begin{abstract}
		We study a quantum network that distributes entangled quantum states to multiple sets of users that are connected to the network. Each user is connected to a switch of the network via a link. All the links of the network generate bipartite Bell-state entangled states in each time-slot with certain probabilities, and each end node stores one qubit of the entanglement generated by the link.  To create shared entanglements for a set of users, measurement operations are performed on qubits of link-level entanglements on a set of related links, and these operations are probabilistic in nature and are successful with certain probabilities. Requests arrive to the system seeking shared entanglements for different sets of users. Each request is for the creation of shared entanglements for a fixed set of users using link-level entanglements on a fixed set of links. Requests are processed according to First-Come-First-Served service discipline and unserved requests are stored in buffers. Once a request is selected for service, measurement operations are performed on qubits of link-level entanglements on related links to create a shared entanglement. For given set of request arrival rates and link-level entanglement generation rates, we obtain necessary conditions for the stability of queues of requests. In each time-slot, the scheduler has to schedule entanglement swapping operations for different sets of users to stabilize the network. Next, we propose a Max-Weight scheduling policy and show that this policy stabilizes the network for all feasible arrival rates. We also provide numerical results to support our analysis. The analysis of a single quantum switch that creates multipartite entanglements for different sets of users is a special case of our work.
	\end{abstract}
	
	%
	%

	\maketitle
	\section{Introduction}
Quantum entanglement is a key component of quantum information systems. It enables applications in quantum key distribution (QKD)\citep{Bennett2014,Ekert}, quantum sensing\citep{Eldredge} (e.g., multipartite entanglement for quantum metrology, \citep{Giovannetti_2011}, \citep{Xia:2019jil}), and distributed quantum computing\citep{Broadbent}.  These applications motivate the need for a distributed infrastructure (quantum network) that will supply high quality (fidelity) bipartite and multipartite entanglement to end groups of users \citep{Pirandola_2019,Pant,Dahlberg_2019,Van_meter,Bhaskar_2020}.  To this end, several network architectures have been proposed to provide high entanglement rates at high fidelity \citep{lee2020quantum,Ruoyu,Armstrong_2012,Herbauts:13,Hall}. 

In this paper we study a quantum network serving entanglement to $M$ groups of users. Here a quantum network consists of a collection of quantum switches connected to each other through optical links. Users requesting service are connected to the network as end nodes.  They have fixed paths between them (in the case of groups consisting of two users), or a fixed spanning three (in the case of groups of three or more users. Every link in the network generates maximally entangled Bell-pairs between two nodes of the link and each of the two nodes stores one qubit of an entanglement of the link in memories. When enough link-level entanglement is accrued (Bell pairs at the links on a path or tree connecting a group of users),  the switches perform multi-qubit measurements to provide end-to-end entanglement to the user group. If the switch only has to connect two links, it uses Bell-state measurements (BSMs) and when it must connect three or more links, it uses Greenberger-Horne-Zeilinger (GHZ) basis measurements \citep{nielsen00}.

We assume that the groups generate entanglement requests  at different rates. Within each group, requests are processed according to First-Come-First-Serve service discipline and waiting requests are stored in corresponding infinite size buffers

We consider a heterogeneous network, in that different types of requests arrive at different rates, links create link-level entanglements with different probabilities, and the measurement success probabilities at switches can differ across switches. Time is divided into time-slots and each link creates at most one entanglement per time-slot, which decoheres or is assumed to be lost after one time-slot\citep{Boxi_li}. Although the expectation is that eventually quantum networks will include switches with many long coherence time quantum memories, this will not be the case in the near term.  For example, first generation quantum networks are likely to use controllable optical delay line buffers \citep{Burmeister:08} to store single qubitsat a time. Furthermore, such models with small coherence times are useful to study quantum networks that are used for QKD applications \citep{nielsen00,Van_meter}. 

In quantum networks of interest, in each time-slot we should associate links with entanglements to requests that can be processed, under the condition that each link can be used by only one request as each link can have at most one entanglement. Then, what is the capacity region, \ie, the set of arrival rates for which there exists
a stationary distribution for occupancy of queues and the average waiting times of requests are finite at equilibrium under a scheduling policy? Can we design a scheduling policy that stabilizes the network for all the arrival rates that belong to the capacity region? We address these research questions in this paper. First, we will derive necessary conditions on the set of arrival rates to achieve stability of the network. We then propose a Max-Weight scheduling policy based on link-level entanglement generation success probabilities, measurement success probabilities, and the topology of the network. This policy does not depend on the request arrival rates, but does depend on request queue sizes and other network parameters. Finally, we show that our policy stabilizes the quantum network for all the arrival rates that belong to the capacity region.

\textbf{Related work:}
Recent successful experiments on quantum information exchange \citep{Sasaki,Yin} encourage us to develop efficient resource allocation algorithms and their performance analysis that can guide us to implement quantum networks at full-scale in future.

In \citep{Shchukin}, a quantum network with two users connected by a series of repeaters was studied. Their focus was to compute the expected waiting time required to create an end-to-end entanglement across a path with $n$ links, where each link creates link-level entanglement with certain probability and measurement operations are successful probabilistically. They used Markov chain theory to develop a method to compute average waiting times. However, they were able to derive exact expressions for paths with at most four segments.

The analysis of a single quantum switch that is connected to several links was investigated in \citep{Gayane1,Gayane2, Gayane3}. In \citep{Gayane1}, the authors studied the model under the assumption that the switch's objective is to create bipartite and tripartite entanglements and it has memories with capabilities to store one qubit and two qubits per each link. They derived expressions for the capacity or the rate at which end-to-entanglements are created. They proposed policies that perform better than the policy that uses the time-division-multiplexing to create bipartite and tripartite entanglements. The analysis of a switch that generates end-to-end $n$-partite entanglements was studied in \citep{Gayane2}, where $n\geq 2$. Using Lyapunov stability theory of Markov chains, they showed that the switch is stable if and only if the number links $K$ attached to the switch is greater than or equal to $n$. Furthermore, they derived closed-form expressions for the capacity, the rate at which the switch creates end-o-end entanglements. In \citep{Gayane3}, the authors studied the capacity region of a switch creating bipartite entanglements under the assumption that it has memories of finite and infinite buffer sizes to store qubits of links. In these works, end-to-end entanglements were created for every set of users of size $n$ and they did not consider in modeling of the switch the requests arriving for end-to-end entanglements from different applications. In our model, we create end-to-end entanglements for the set of users that requests ask for. Our objective is to create end-to-end entanglements for requests arriving in the system efficiently using network resources.

In \citep{Leandros_maxweight}, a Max-Weight scheduling policy was first introduced for resource allocation in communication networks and it was shown that the proposed policy stabilizes the network for all feasible arrival rates. Later, this policy was adopted for the analysis of a single classical switch in \citep{McKeown}, where they showed that switch is stable for all feasible arrival rates under this policy. The models studied in \citep{Leandros_maxweight,McKeown} have static network topologies. A salient feature of quantum networks is that they are dynamic in nature, due to the fact that several operations in quantum networks are stochastic in nature. Hence, a Max-Weight scheduling policy for quantum networks must be defined by considering dynamic properties of quantum networks. In a different context, a Max-Weight scheduling policy was studied in \citep{Leandros_Varying} for classical networks with varying topologies. We adopt this formulation for analyzing our quantum networks.

\textbf{Our Contributions:} We make the following contributions:\bit
\item We derive necessary conditions on the request arrival rates for achieving the stability of the network under a scheduling policy.
\item We propose a Max-Weight scheduling policy as a function of probability of successful creation of link-level entanglements and measurement operations, and dynamic queue sizes of requests. We prove that this policy stabilizes the network for all feasible arrival rates using Lyapunov stability theory of Markov chains.
\item Finally, we provide numerical results that corroborate our analysis.
\eit

The rest of the paper is organized as follows.
In Section~\ref{sec:model}, we introduce the model and demonstrate challenges in analyzing our quantum networks. We then give notation and some preliminary results in Section~\ref{sec:preliminary}, where we also define our Max-Weight scheduling policy. In Section~\ref{sec:results}, we give necessary conditions on the request arrival rates for the stability of the network and provide main results along with some proofs. We provide numerical results on the stability of quantum networks under the Max-Weight scheduling policy in Section~\ref{sec:numerics}. Finally, we give concluding remarks in Section~\ref{sec:conclusions}. We give proofs of some supporting results in Appendix.                                               
	\section{System Model}
	\label{sec:model}
We study a quantum network that creates shared entangled quantum states for several sets of users. We model a quantum network as a graph $G = (V,L)$. Here $V = V_n \cup V_u$ where $V_n$ is the set of network nodes and $V_u$ is a set of $N$ user nodes. Here $L$ is the set of links connecting users and network nodes to network nodes, $L \subset V \times V_n$. Time is assumed to be divided into fixed length time-slots. During each time-slot, we assume each link attempts to generate a Bell-pair entanglement between the two nodes of the link and each succeeds with probability $p_i$, $i\in L$. Each node of a link can store up to one qubit of the Bell-pair entanglement generated by the link. This qubit must be used to perform a  measurement by the end of the time-slot or it is lost due to decoherence. Network nodes can perform measurements on groups of two or more successfully created neighboring link entanglements.  

We assume that the network provides shared entangled states for $M$ types of requests. Further, a type $i$ request seeks creation of shared entanglements for a set of users denoted by $U_i \in V_u$ using network nodes and links in a subgraph of $G$, $\mT_i = (X_i,L_i)$, that connects the users in $U_i$.  When $U_i = \{u,v\}$, $\mT_i$ is a path within $G$ connecting $u$ and $v$; when $|U_i| > 2$, $\mT_i$ is a tree within $G$ connecting users in $U_i$.    
	For a type $i$ request, we assume that all involved entanglement swapping operations succeed with probability $q_i = \prod_{j\in X_i \setminus U_i}q_{ij}$ where  $q_{ij}$ denotes the measurement success probability at network node $j\in V_n$. 

	During each time-slot, type $i$ requests arrive to the system according to a stationary process $\{A_{i}(n)\}$, where, $A_{i}(n)$ denotes the number of type~$i$ requests arrived in time-slot $n$ and the average number of requests arrived per time-slot is denoted by $\lambda_{i}$. For each application, the system has a buffer with infinite memory to store unserved requests, and requests are served according to the First-Come-First-Served (FCFS) service discipline.
	
\begin{figure}
		\centering
		\includegraphics[width=0.65\linewidth]{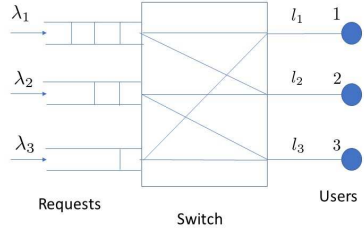}
		\caption{Switch creating bipartite entanglements}
		\label{fig:switch1}
	\end{figure}

	
	A single quantum switch is an example for a simple quantum network. In Figure~\ref{fig:switch1}, we show a quantum switch that connects to three users, where user $i$ is connected to the switch via link $l_i$. There are three types of requests arriving in the system, each type of request seeks a creation of shared entanglement for a pair of users. For type $i$ requests, $\lambda_i$ denotes the average number of requests arriving in each time-slot. 
	A type~$i$ request seeks an end-to-end entanglement for user pair $U_i$ using link level entanglements on links $L_i$. From Figure~\ref{fig:switch1}, the set of user pairs and link pairs for different sets of requests are: $U_1=\{1,2\}$, $U_2=\{2,3\}$, $U_3=\{1,3\}$, $L_1=\{l_1,l_2\}$, $L_2=\{l_2,l_3\}$, and $L_3=\{l_1,l_3\}$.
	
\begin{figure}
		\centering
		\includegraphics[width=0.95\linewidth]{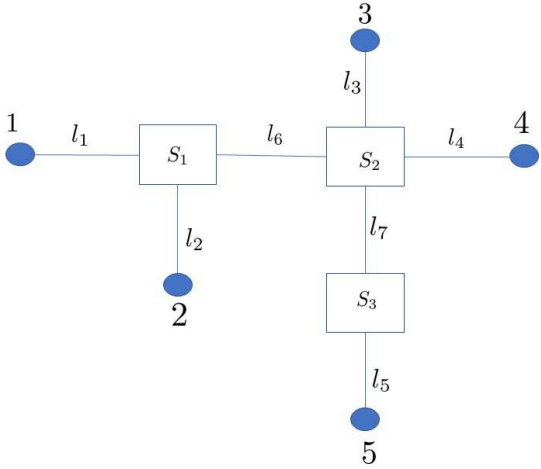}
		\caption{A quantum network with five users} 
		\label{fig:switch_network}
	\end{figure}

In Figure~\ref{fig:switch_network}, we show a quantum network connected to five users. This network has three switches $\{S_1,\cdots,S_3\}$ and seven links $\{l_1,\cdots,l_7\}$, where user $i$ is connected to the network via link $l_i$. Suppose that there are two types of requests with $U_1=\{1,2,3\}$, $U_2=\{4,5\}$, $L_1=\{l_1,l_2,l_3,l_6\}$, and $L_2=\{l_4,l_5,l_7\}$. Now the objective is to serve requests using qubits of associated links for each type of requests.	

	Next we use the following simple model to illustrate the key questions that we plan to address.
	Consider a single switch as in Figure~\ref{fig:switch1} connected to three users via three links. The objective is to create end-to-end entanglements for user pairs. Suppose that all the three links have qubits in a given time-slot. Then the key challenge is, how does the switch decide to serve requests for user pairs in order to stabilize the network? Since each link has only one qubit, out of the three user pairs, a request of only one user pair can be served in the given time-slot as each link can be paired-up with only one other link. In general, how do you select link-pairs in each time-slot such that each link is paired-up with only one other link? More importantly, the switch's goal is to schedule entanglement swappings for link-pairs in each time-slot so that the queues of requests are stable, in that the average number of stored requests for end-to-end entanglements for each link-pair does not scale-up with time. We need to design a policy that governs the switch to make scheduling decisions in each time-slot such that the switch is stable. Similarly, is there any scheduling policy that stabilizes a given quantum network?

	In each time-slot, suppose that the quantum nodes provide information regarding the availability of Bell-pair entanglements of attached links and queue sizes of requests to a centralized scheduler and further assume that the scheduler maintains queues for each of the $M$ types of requests. The goal of the scheduler is to select a set of requests that can be served in the given time-slot which then provides the details of the scheduling decision to the quantum nodes.
	We would like to answer the following question. For the system with $K$ links and $N$ users, for given $\mf{p}=[p_1,\cdots,p_K]$ and $\mf{q}=[q_1,\cdots,q_M]$, what is the capacity region of request rates? Here, the capacity region is the set of request rates $\bm{\lambda}=[\lambda_1, \cdots,\lambda_M]$ for which there exists a scheduling policy that achieves the stability of the system. 

	In the next section, we define our Max-Weight scheduling policy and give some preliminary results.
	\section{Notation and Preliminary Results}
	\label{sec:preliminary}
	
	Throughout the paper we use bold-faced letters for vectors.
	Next, we introduce some terminology of the paper below:
	\begin{align}
		Q_{i}(n)&: \text{ queue size of type~$i$ requests } \text{at the beginning of time-slot } n\nonumber\\
		A_{i}(n)&: \text{ number of type $i$ requests that arrive } \text{in time-slot }  n\nonumber\\
		T_{i}(n)&:  \text{ }T_{i}(n)=1 \text{  if link } i \text{ entanglement is successful in time-slot } n; \text{ otherwise }T_i(n)=0.\nonumber
	\end{align}

	In each time-slot, since each link has at most one entanglement, it can be used to serve at most one request during that time-slot. We need to select a set of requests for service such that each link can be a member of at most one group of links associated with those requests. This is achieved using a notion called matching, that is defined below. 
	
	\begin{defn}{Matching:}
		\label{defn:matching}
		We call $\bm{\pi}=[\pi_{1},\cdots,\pi_M]$ a matching if $\pi_{i}\in\{0,1\}$ and for each link $j$ ($1\leq j\leq K$), we have
		\beq
		\label{eq:matching}
		\sum_{i=1,j\in L_i}^M\pi_{i}\leq 1.
		\eeq
		Furthermore, if $\pi_r=0$ ($1\leq r\leq K$) in $\bm{\pi}$, then the vector $\bm{\pi}^*$ obtained from $\bm{\pi}$ by replacing the $r^{\text{th}}$ element $\pi_r=0$ with $\pi_r=1$, violates condition \eqref{eq:matching}. 
	\end{defn}
	Let $\mc{M}$ be the set of all matchings defined as
	\beq
	\mc{M}\triangleq\{\bm{\pi}: \bm{\pi}\text{ is a matching}\}.
	\eeq.

	If the scheduler selects a matching $\bm{\pi}=[\pi_{i},1\leq i\leq K]$ to serve requests in a time-slot, then an entanglement swapping operation is performed on qubits of links $L_i$ if $\pi_{i}=1$. 
	%
	In time-slot $n$, based on $\mf{Q}(n)=(Q_{i}(n))$ and $\mf{T}(n)=(T_i(n))$, the scheduler selects a matching according to a scheduling policy. Suppose that $\mf{W}(n)$ is the matching used in time-slot $n$, then let $u_i(\mf{Q}(n),\mf{T}(n),\mf{W}(n))$ denote the probability that a type~$i$ request is successfully served conditioned on the event that the matching $\mf{W}(n)$ is used to schedule requests for service. Let $\indic{A}$ be the indicator function of the event $A$. Then we write
	\beq
	u_i(\mf{Q}(n),\mf{T}(n),\mf{W}(n))=q_i\indic{W_i(n)>0}\indic{Q_i(n)>0}\indic{T_j(n)>0,\forall j\in L_i},
	\eeq
	where $q_i$ denotes the joint success probability of all the measurement operations performed on qubits of links $L_i$ to create a shared entanglement.
	Next, we define the Max-Weight scheduling policy of interest below.
	\begin{defn}{Max-Weight Scheduling:}
		In time-slot $n$, under the Max-Weight scheduling policy, the scheduler selects the matching $\mf{W}(n)$ computed as follows:
		\beq
		\label{eq:max_weight}
		\mf{W}(n)=\arg\max_{\bm{\pi}\in\mc{M}}\sum_{i=1}^Mu_i(\mf{Q}(n),\mf{T}(n),\bm{\pi})Q_{i}(n).
		\eeq
	\end{defn}
	

	Next, we write the dynamics of the process $\{\mf{Q}(n)\}$. Let $Z_i(n)\in\{0,1\}\}$ denote whether all the measurement operations required to create a shared entanglement using qubits of links $L_i$ in time-slot $n$ succeed or not. Here $Z_i(n)=1$ represents that all measurement operations succeed; otherwise, $Z_i(n)=0$.
	We denote the measurement success probability as $\mb{P}(Z_i(n)=1)=q_i$. Now denote by $D_i(n)\in\{0,1\}$ the number of requests served for application $i$ in time-slot $n$. We write
	\beq
	D_i(n)=Z_i(n)\indic{W_i(n)>0}\indic{Q_i(n)>0}\indic{T_j(n)>0,\forall j\in L_i}. 
	\eeq
	The evolution of the process $\{\mf{Q}(n)\}$	is described using the following relation
	\beq
	\label{eq:queue_dynamics}
	\mf{Q}(n+1)=\mf{Q}(n)-\mf{D}(n)+\mf{A}(n).
	\eeq
	Note that the arrived requests $\mf{A}(n)$	are not used in evaluating the served requests $\mf{D}(n)$, but they are used in the computation of $\mf{D}(n+1)$. If $\mf{D}(n)=\mf{0}$, then it means that no request is served in time-slot $n$.
	
	Let the number of served requests in each time-slot be denoted by $\bm{\sigma}=(\sigma_{i})$, where $\sigma_{i}\in\{0,1\}$ is the number of type~$i$ served requests and $\bm{\sigma}$ belongs to the set $\mc{D}$ defined by
	\beq
	\mc{D}\triangleq \{\mf{a}=(a_{i}): a_{i}\in\{0,1\},\sum_{i=1,j\in L_i}^Ma_{i}\leq 1,\,\forall j\}.
	\eeq
	Note that the zero vector $\mf{0}$ is also an element of the set $\mc{D}$.

	%


	Now we obtain a necessary condition on the arrival rates that must be satisfied for the system to be stable. For type~$i$ requests, since a link $j\in L_i$ generates an entanglement with probability $p_j$ and the joint success probability of the involved measurement operations is $q_i$, it follows that
	\beq
	\lambda_{i}<q_i\prod_{j\in L_i}p_j.
	\eeq
	For a link $r\in L_i$ we can write
	\beq
	\lambda_{i}<p_rq_i\prod_{j\in L_i,j\neq r}p_j.
	\eeq
	Here, we interpret $\frac{\lambda_{i}}{q_i\prod_{j\in L_i,j\neq r}p_j}$ as the average number of arrived type~$i$ requests per time-slot, conditioned on the event that all the links $L_i$ except link $r\in L_i$ have entanglements and measurement operations performed on qubits of links $L_i$ are successful. Since each link can belong to at most one group of links that can serve a request, we must have
	\beq
	\sum_{i=1,r\in L_i}^M\frac{\lambda_{i}}{q_i\prod_{j\in L_i,j\neq r}p_j}<p_r.
	\eeq
	Therefore, a necessary condition is that for each link $r$,
	\beq
	\label{eq:necessary}
	\sum_{i=1,r\in L_i}^M\frac{\lambda_{i}}{q_i\prod_{j\in L_i}p_j}<1.
	\eeq
	
	Let $\Lambda$ be defined as
	as
	\beq
	\label{eq:nec1}
	\Lambda\triangleq \left\{\bm{\lambda}: \sum_{i=1,r\in L_i}^M\frac{\lambda_{i}}{q_i\prod_{j\in L_i}p_j}<1, \forall\, 1\leq r\leq K \right\}.
	\eeq
	
	Now we investigate whether there exists a scheduling policy that achieves the stability of the system for each $\bm{\lambda}\in \Lambda$. To this end, we present the following example. 
	\begin{ex}
		Consider a quantum switch as in Figure~\ref{fig:switch1} with three links ($K=3$) that provides service to three types of applications ($M=3$) requesting end-to-end bipartite entanglements using different sets of link-pairs. Assume that $U_1=\{1,2\}$, $U_2=\{2,3\}$, and $U_3=\{1,3\}$. Suppose that $\lambda_{1}=\lambda_{2}=\lambda_{3}=a$ and $p_i=1$ for $1\leq i\leq 3$. Furthermore, let $q_i=1$ for all $i$, $1\leq i\leq 3$. For this model, it is easy to check that only one request can be served in each time-slot under the assumption that qubits decohere after one time-slot. Because of symmetry, each type of requests gets a service rate of $\frac{1}{3}$. Then if $a\geq \frac{1}{3}$, the switch is unstable.  
		From the condition $\bm{\lambda}\in\Lambda$,
		we obtain $a\leq \frac{1}{2}$. It shows that the switch is unstable if $\frac{1}{3}\leq a \leq \frac{1}{2}$. Hence, there exists a $\bm{\lambda}\in\Lambda$ for which there is no scheduling policy for which the switch is stable.
	\end{ex}

	From the above example, it is clear that the switch cannot serve more than one request in one time-slot. Since each link has at most one entanglement in each time-slot, a link can be paired-up with only one other link. The condition \eqref{eq:necessary} is derived by considering the traffic arriving at a single link. However, the necessary condition should also be derived based on how many requests of different types can be served simultaneously in one time-slot using qubits of several links. The above example shows that the capacity region is smaller than $\Lambda$ due to the fact that some sets of requests cannot be served simultaneously in one time-slot.

	\begin{remark}
		In classical networking, a switch forwards requests from input ports to output ports. Let $\lambda_{ij}$ denotes the average number of arrived requests at the input port $i$ for the output port $j$ per time-slot. In each time-slot, at most one request can be forwarded from each input port to only one output port, and also, each output port receives at most one request from only one input port. Define $\Lambda'$ as
		\beq
		\Lambda'=\{\mf{a}=[a_{ij}]:\sum_{j}a_{ij}\leq 1\text{ and }\sum_{l}a_{lm}\leq 1,\, \forall\,i,m\}.
		\eeq
		Let $\mc{M}'$ be the set of classical matchings defined as
		\beq
		\mc{M}'\triangleq\{\bm{\pi}=[\pi_{ij}]:\sum_{j}\pi_{ij}= 1\text{ and }\sum_{l}\pi_{lm}= 1,\, \forall\,i,m\}.
		\eeq
		In \citep{McKeown}, it was shown that if the switch selects the matching $W(n)$ computed according to the following Max-Weight scheduling policy,
		\beq
		\label{eq:max_weight2}
		\mf{W}(n)=\arg\max_{\pi\in\mc{M}'}\sum_{ij}\pi_{ij}Q_{ij}(n),
		\eeq
		then the switch is stable if $\bm{\lambda}$ lies inside $\Lambda'$. Furthermore, if $\bm{\lambda}\notin \Lambda'$, then no scheduling policy achieves the stability of the switch. We can view the quantum switch as the device with $M$ input ports and $K$ output ports, where each input port is associated with an application and each output port is associated with a link. In each time-slot, input port $i$ is either matched to output ports $L_i$ or not matched to any output port. Furthermore, each output port is matched to  at most one input port. If the input port $i$ is matched to output ports, then it means that the switch attempts to serve a type~$i$ request. 	\end{remark}
	
	In the next section, we will derive a final necessary condition on $\bm{\lambda}$ for the stability of the system and show that the proposed Max-Weight scheduling policy achieves the stability of the system. 
	
	\section{Main Results}
	\label{sec:results}
	In this section, we present the main results of the paper. The process $\{(\mf{Q}(n))\}$, where $\mf{Q}(n)=(Q_{i}(n),1\leq i\leq M)$ is a Markov chain. Our goal is to find necessary conditions on system parameters to achieve stability of the system and also show that the Max-Weight scheduling policy achieves the stability of the system.


	We begin by obtaining conditions on $\bm{\lambda}$ that are required to achieve the stability of the system. From \eqref{eq:queue_dynamics}, we can write
	\beq
	\sum_{j=1}^n\mf{Q}(j+1)=\sum_{j=1}^n\mf{Q}(j)-\sum_{j=1}^n\mf{D}(j)+\sum_{j=1}^n\mf{A}(j).
	\eeq
	Hence, we obtain
	\beq
	\mf{Q}(n+1)=\mf{Q}(1)-\sum_{j=1}^n\mf{D}(j)+\sum_{j=1}^n\mf{A}(j).
	\eeq
	Here, $\sum_{j=1}^nD_{i}(j)$ denotes the cumulative number of type~$i$ served requests up to time-slot $n$. Similarly, 
	$\sum_{j=1}^nA_{k}(j)$ denotes the cumulative number of arrived requests of type~$k$ up to time-slot $n$. 
	We can also write
	\beq
	\sum_{j=1}^n\mf{D}(j)=\sum_{\bm{\sigma}\in\mc{D}}c_{\bm{\sigma}}(n)\bm{\sigma}\text{ and }\sum_{\bm{\sigma}\in\mc{D}}c_{\bm{\sigma}}(n)=n,
	\eeq
	where $c_{\bm{\sigma}}(n)$ defined as
	\beq
	c_{\bm{\sigma}}(n)=\sum_{i=1}^n\indic{\mf{D}(i)=\bm{\sigma}},
	\eeq
	denotes the cumulative number of times the vector $\bm{\sigma}$ is served up to time-slot $n$. 
	In the following theorem, we show a relationship between $\bm{\lambda}$ and $\bm{\sigma}$ if the system is stable under a scheduling policy.

	Let $\mc{P}_{\mc{D}}$ be the set of probability distributions on the set $\mc{D}$ defined as
	\beq
	\mc{P}_{\mc{D}}\triangleq\{\mf{a}=\{a_{\bm{\sigma}},\bm{\sigma}\in\mc{D}\}:a_{\bm{\sigma}}\in\mb{R}_+,\,\sum_{\bm{\sigma}\in\mc{D}}a_{\bm{\sigma}}=1\},
	\eeq
	where, $\mb{R}_+$ denotes the set of non-negative real numbers.

	\begin{theorem}
		\label{thm:stable}
		For the arrival rate matrix $\bm{\lambda}$ and under the assumption $\mb{E}[\mf{Q}(1)]<\infty$, 
		if the system is stable under a scheduling policy then there exists $\mf{c}^*=(c^*_{\bm{\sigma}},\bm{\sigma}\in\mc{D})\in\mc{P}_{\mc{D}}$ such that
		\beq
		\label{eq:serv_vecs}
		\bm{\lambda}=\sum_{\bm{\sigma}\in\mc{D}}c_{\bm{\sigma}}^*\bm{\sigma},
		\eeq
		where $c_{\bm{\sigma}}^*$ denotes the probability with which $\bm{\sigma}$ is served in a time-slot at equilibrium.
	\end{theorem}
	\begin{proof}
		The proof is given in Appendix~\ref{app:stable}.
		
	\end{proof}
	
	The probability distribution $\mf{c}^*$ is defined in terms of the stationary distribution of $\mf{Q}(n)$ and the distribution of $\mf{T}(n)$.

	A vector $\bm{\sigma}$ is served in a time-slot for some set of values of $\mf{Q}(n)$, $\mf{T}(n)$, and $\mf{W}(n)$. Let the random variables with $n=\infty$ correspond to the variables with stationary distributions. Then from the definition of $\mf{D}(n)$,  \eqref{eq:serv_vecs} is equivalent to the following equation
	\begin{multline}
		\label{eq:nec4}
	\bm{\lambda}=\sum_{\mf{a}\in\{0,1\}^K}\mb{P}(\mf{T}(\infty)=\mf{a})\mb{P}(\mf{Q}(\infty)\neq 0)\\
	\times\sum_{\mf{b}\in\mb{Z}_+^M,\mf{b}\neq 0}\mb{P}(\mf{Q}(\infty)=b\given \mf{Q}(\infty)\neq 0)\sum_{\bm{\pi}\in\mc{M}}\mb{P}\Big(\mf{W}(\infty)=\bm{\pi}\given \mf{Q}(\infty)=\mf{b},\mf{T}(\infty)=\mf{a}\Big)\mf{u}(\mf{b},\mf{a},\bm{\pi}),
	\end{multline}
where $\mb{Z}_+$ denotes the set of non-negative integers.

	 From \eqref{eq:nec4}, if the system is stable, then we obtain the following result.
	\begin{corollary}
		\label{cor:stable}
		If there exists a scheduling policy that stabilizes the system, then  
		\beq
		\label{eq:nec5}
			\bm{\lambda}=\sum_{\mf{a}\in\{0,1\}^K}\mb{P}(\mf{T}(\infty)=\mf{a})
			\sum_{\mf{b}\in\mb{Z}_+^M,\mf{b}\neq 0}\sum_{\bm{\pi}\in\mc{M}}
			r(\mf{b},\mf{a},\bm{\pi})\mf{u}(\mf{b},\mf{a},\bm{\pi}),
		\eeq
		where $r(\mf{b},\mf{a},\bm{\pi})$ denotes the stationary probability that the vector of queue sizes is $\mf{b}$ and the selected matching is $\bm{\pi}$ conditioned on the event that $\mf{T}(n)$ is equal to $\mf{a}$, and $\sum_{\mf{b}\in\mb{Z}_+^M,\mf{b}\neq 0}\sum_{\bm{\pi}\in\mc{M}}
		r(\mf{b},\mf{a},\bm{\pi})<1$ for all $\mf{a}$.
	\end{corollary}
	\begin{proof}
		Since the system is stable, there exists a non-zero probability that the stationary queue sizes of requests satisfies $\bm{Q}(\infty)=\bm{0}$. As a result, $\sum_{\mf{b}\in\mb{Z}_+^M,\mf{b}\neq 0}\sum_{\bm{\pi}\in\mc{M}}
		r(\mf{b},\mf{a},\bm{\pi})<1$.
	\end{proof}

	%
	%
	%
	%
	%
	%
	%
	%
	
	Next, we use Corollary~\ref{cor:stable} to conditions on $\bm{\lambda}$ for achieving stability of the network.
	Define $\Lambda_Q$ to be
	\begin{multline}
	\Lambda_Q\triangleq\Big\{\bm{\lambda}\in\Lambda:\exists\,\,\mf{v}= \{v(\mf{b},\mf{a},\bm{\pi})\in\mf{R}_+\}\text{ such that }\\
	\bm{\lambda}=\sum_{\mf{a}\in\{0,1\}^K}\mb{P}(\mf{T}(\infty)=\mf{a})
	\sum_{\mf{b}\in\mb{Z}_+^M,\mf{b}\neq 0}\sum_{\bm{\pi}\in\mc{M}}
	v(\mf{b},\mf{a},\bm{\pi})\mf{u}(\mf{b},\mf{a},\bm{\pi})\\
	\text{ and }
	\sum_{\mf{b}\in\mb{Z}_+^M,\mf{b}\neq 0}\sum_{\bm{\pi}\in\mc{M}}
	v(\mf{b},\mf{a},\bm{\pi})<1\Big\}.
	\end{multline}
	\begin{theorem}
		If $\bm{\lambda}\notin\Lambda_{Q}$, then no scheduling policy stabilizes the system.
	\end{theorem}
	\begin{proof}
		Suppose there exists a scheduling policy that stabilizes the system. From Corollary~\ref{cor:stable}, using the stationary distribution of the system, we obtain \eqref{eq:nec5}.  This is a contradiction. Hence, no policy can stabilize the system if $\bm{\lambda}\notin\Lambda_{Q}$. 
	\end{proof}
	
	It is of interest to check whether $\Lambda=\Lambda_Q$ or not. In the next example, we present a $\bm{\lambda}\in\Lambda$ that does not belong to $\Lambda_Q$. Hence, $\Lambda_Q\subset\Lambda$.
	\begin{ex}
		Consider a single switch that creates end-to-end entanglements for user-pairs with $K=4$, $M=6$, $p_i=1$ for $1\leq i\leq K$, and $q_j=1$ for $1\le j\leq M$. The set of user-pairs are: $U_1=\{1,2\}$,  $U_2=\{1,3\}$, $U_3=\{1,4\}$, $U_4=\{2,3\}$,$U_5=\{2,4\}$, and $U_6=\{3,4\}$. The arrival rate vector is $\bm{\lambda}=[0.3\,\, 0.3\, \,0.3 \,\,0.2\, \,0.45\,\, 0.2]$. It can be verified that $\bm{\lambda}\in\Lambda$.   A partitioning of $\bm{\lambda}$ is given by
		\begin{multline}
			\bm{\lambda}=0.2\begin{bmatrix}
				1 & 0 &  0 &  0&0&1
			\end{bmatrix}+0.2\begin{bmatrix}
			0 & 0 &  1 &  1&0&0
		\end{bmatrix}+0.3\begin{bmatrix}
		0 & 1 &  0 &  0&1&0
	\end{bmatrix}\\
			+0.1\begin{bmatrix}
				1 & 0 &  0 &  0&0&0
			\end{bmatrix}
			+0.1\begin{bmatrix}
				0 & 0 &  1 &  0&0&0
			\end{bmatrix}+0.15\begin{bmatrix}
			0 & 0 &  0 &  0&1&0
		\end{bmatrix}.
			\label{eq:partition_ex}
		\end{multline}
		In the above partition, each vector on the right side of \eqref{eq:partition_ex} corresponds to a matching in $\mc{M}$. The sum of the coefficients in the partition of $\bm{\lambda}$ given in \eqref{eq:partition_ex} is equal to $0.2+0.2+0.3+0.1+0.1+0.15=1.05$. 
		Furthermore, it is not possible to partition $\bm{\lambda}$ such that the sum of coefficients in the partition is less than one. 
		As a result, we conclude that $\bm{\lambda}\notin\Lambda_Q$. Hence, if the requests arrive in the system such that the average number of arrived requests per time-slot is $\bm{\lambda}$, then the switch is unstable under any scheduling policy.
	\end{ex}


	Next, we establish a preliminary result before presenting the main result on the stability of the network under the Max-Weight scheduling policy. Let $\mf{c}\cdot\mf{d}$ be the inner product of two vectors $\mf{c}$ and $\mf{d}$.
	\begin{lemma}
	\label{thm:gap_bound_lemma}
	If $\bm{\lambda}\in\Lambda_Q$, then
	\beq
	\bm{\lambda}\cdot\mf{Q}(n)-\sum_{\mf{a}\in\{0,1\}^K}\mb{P}(\mf{T}(n)=\mf{a})\max_{\bm{\pi}\in\mc{M}}\left(\sum_{i=1}^M u_i(\mf{Q}(n),\mf{a},\bm{\pi}))Q_i(n)\right)<-\epsilon \norm{\mf{Q}(n)},
	\eeq
	where $\norm{\mf{Q}(n)}=\sqrt{\sum_{i=1}^MQ_i^2(n)}$.
	\end{lemma}
	\begin{proof}
	Since $\bm{\lambda}\in\Lambda_Q$, there exists $\mf{v}=\{v(\mf{b},\mf{a},\bm{\pi})\in\mb{R}_+\}$ such that
	\beq
	    \bm{\lambda}=\sum_{\mf{a}\in\{0,1\}^K}\mb{P}(\mf{T}(n)=\mf{a})
	\sum_{\mf{b}\in\mb{Z}_+^M,\mf{b}\neq 0}\sum_{\bm{\pi}\in\mc{M}}
	v(\mf{b},\mf{a},\bm{\pi})\mf{u}(\mf{b},\mf{a},\bm{\pi}).
	\eeq
	Hence, we can write
	\begin{multline}
	\label{eq:gap_bound}
		\bm{\lambda}\cdot\mf{Q}(n)-\sum_{\mf{a}\in\{0,1\}^K}\mb{P}(\mf{T}(n)=\mf{a})\max_{\bm{\pi}\in\mc{M}}\left(\sum_{i=1}^M u_i(\mf{Q}(n),\mf{a},\bm{\pi}))Q_i(n)\right)\\
		=\sum_{\mf{a}\in\{0,1\}^K}\mb{P}(\mf{T}(n)=\mf{a})
	\sum_{\mf{b}\in\mb{Z}_+^M,\mf{b}\neq 0}\sum_{\bm{\pi}\in\mc{M}}
	v(\mf{b},\mf{a},\bm{\pi})\sum_{i=1}^Mu_i(\mf{b},\mf{a},\bm{\pi})Q_i(n)\\
	-\sum_{\mf{a}\in\{0,1\}^K}\mb{P}(\mf{T}(n)=\mf{a})\max_{\bm{\pi}\in\mc{M}}\left(\sum_{i=1}^M u_i(\mf{Q}(n),\mf{a},\bm{\pi}))Q_i(n)\right).
	\end{multline}
	We use the following bound to simplify the above equation
	\begin{align}
	\label{eq:ui_bound}
	u_i(\mf{b},\mf{a},\bm{\pi})Q_i(n)&=	q_i\indic{\pi_i>0}\indic{b_i>0}\indic{a_j>0,\forall j\in L_i}Q_i(n)\nonumber\\
	&\leq q_i\indic{\pi_i>0}\indic{a_j>0,\forall j\in L_i}Q_i(n)\nonumber\\
	&= q_i\indic{\pi_i>0}\indic{Q_i(n)>0}\indic{a_j>0,\forall j\in L_i}Q_i(n)\nonumber\\
	&=u_i(\mf{Q}(n),\mf{a},\bm{\pi})Q_i(n)
	\end{align}
	By substituting \eqref{eq:ui_bound} in \eqref{eq:gap_bound}, we obtain
	\begin{multline}
		\bm{\lambda}\cdot\mf{Q}(n)-\sum_{\mf{a}\in\{0,1\}^K}\mb{P}(\mf{T}(n)=\mf{a})\max_{\bm{\pi}\in\mc{M}}\left(\sum_{i=1}^M u_i(\mf{Q}(n),\mf{a},\bm{\pi}))Q_i(n)\right)\\
		=\sum_{\mf{a}\in\{0,1\}^K}\mb{P}(\mf{T}(n)=\mf{a})
	\sum_{\mf{b}\in\mb{Z}_+^M,\mf{b}\neq 0}\sum_{\bm{\pi}\in\mc{M}}
	v(\mf{b},\mf{a},\bm{\pi})\sum_{i=1}^Mu_i(\mf{Q}(n),\mf{a},\bm{\pi})Q_i(n)\\
	-\sum_{\mf{a}\in\{0,1\}^K}\mb{P}(\mf{T}(n)=\mf{a})\max_{\bm{\pi}\in\mc{M}}\left(\sum_{i=1}^M u_i(\mf{Q}(n),\mf{a},\bm{\pi}))Q_i(n)\right).
	\end{multline}
	We can write the above equation as
	\begin{multline}
		\bm{\lambda}\cdot\mf{Q}(n)-\sum_{\mf{a}\in\{0,1\}^K}\mb{P}(\mf{T}(n)=\mf{a})\max_{\bm{\pi}\in\mc{M}}\left(\sum_{i=1}^M u_i(\mf{Q}(n),\mf{a},\bm{\pi}))Q_i(n)\right)\\
		=\sum_{\mf{a}\in\{0,1\}^K}\mb{P}(\mf{T}(\infty)=\mf{a})
	\sum_{\mf{b}\in\mb{Z}_+^M,\mf{b}\neq 0}\sum_{\bm{\pi}\in\mc{M}}
	v(\mf{b},\mf{a},\bm{\pi})\\
	\times \left(\sum_{i=1}^Mu_i(\mf{Q}(n),\mf{a},\bm{\pi})Q_i(n)-\max_{\bm{\pi}\in\mc{M}}\sum_{i=1}^Mu_i(\mf{Q}(n),\mf{a},\bm{\pi})Q_i(n)\right)\\
	-\sum_{\mf{a}\in\{0,1\}^K}\mb{P}(\mf{T}(n)=\mf{a})\left(1-\sum_{\mf{b}\in\mb{Z}_+^M,\mf{b}\neq 0}\sum_{\bm{\pi}\in\mc{M}}
	v(\mf{b},\mf{a},\bm{\pi})\right)\max_{\bm{\pi}\in\mc{M}}\left(\sum_{i=1}^M u_i(\mf{Q}(n),\mf{a},\bm{\pi}))Q_i(n)\right).
	\end{multline}
	Since the first term on the right side of the above equation is negative, we obtain
	\beq
		\bm{\lambda}\cdot\mf{Q}(n)-\sum_{\mf{a}\in\{0,1\}^K}\mb{P}(\mf{T}(n)=\mf{a})\max_{\bm{\pi}\in\mc{M}}\left(\sum_{i=1}^M u_i(\mf{Q}(n),\mf{a},\bm{\pi}))Q_i(n)\right)\nonumber
		\eeq
		\begin{align}
	&	\leq
	-\sum_{\mf{a}\in\{0,1\}^K}\mb{P}(\mf{T}(n)=\mf{a})\left(1-\sum_{\mf{b}\in\mb{Z}_+^M,\mf{b}\neq 0}\sum_{\bm{\pi}\in\mc{M}}
	v(\mf{b},\mf{a},\bm{\pi})\right)\max_{\bm{\pi}\in\mc{M}}\left(\sum_{i=1}^M u_i(\mf{Q}(n),\mf{a},\bm{\pi}))Q_i(n)\right)\nonumber\\
&	\leq
	-m\sum_{\mf{a}\in\{0,1\}^K}\mb{P}(\mf{T}(n)=\mf{a})\max_{\bm{\pi}\in\mc{M}}\left(\sum_{i=1}^M u_i(\mf{Q}(n),\mf{a},\bm{\pi}))Q_i(n)\right),\label{eq:gap_bound2}
	\end{align}
	where $m=\min_{\mf{a}}\left(1-\sum_{\mf{b}\in\mb{Z}_+^M,\mf{b}\neq 0}\sum_{\bm{\pi}\in\mc{M}}
	v(\mf{b},\mf{a},\bm{\pi})\right)>0$.
	
	Suppose that $i^*=\arg\max_i Q_i(n)$. Then we can write \eqref{eq:gap_bound2} as
	\beq
		\bm{\lambda}\cdot\mf{Q}(n)-\sum_{\mf{a}\in\{0,1\}^K}\mb{P}(\mf{T}(n)=\mf{a})\max_{\bm{\pi}\in\mc{M}}\left(\sum_{i=1}^M u_i(\mf{Q}(n),\mf{a},\bm{\pi})Q_i(n)\right)\nonumber
		\eeq
			\begin{align}
		&\leq
	-m\sum_{\mf{a}\in\{0,1\}^K}\mb{P}(\mf{T}(n)=\mf{a})\max_{\bm{\pi}\in\mc{M}}\left( u_{i^*}(\mf{Q}(n),\mf{a},\bm{\pi})Q_{i^*}(n)\right)\nonumber\\
	&\leq
	-m\,\max_{\mf{a}\in\{0,1\}^K}\left(\mb{P}(\mf{T}(n)=\mf{a})\max_{\bm{\pi}\in\mc{M}}\left( u_{i^*}(\mf{Q}(n),\mf{a},\bm{\pi})Q_{i^*}(n)\right)\right)\nonumber\\
		&\leq
	-m\, \left(\max_{\mf{a}\in\{0,1\}^K}\max_{\bm{\pi}\in\mc{M}}\mb{P}(\mf{T}(n)=\mf{a})\left( u_{i^*}(\mf{Q}(n),\mf{a},\bm{\pi})\right)\right)Q_{i^*}(n)\nonumber\\
	&=
	-m\, \left(\max_{\mf{a}\in\{0,1\}^K}\max_{\bm{\pi}\in\mc{M}}\mb{P}(\mf{T}(n)=\mf{a})\left( q_{i^*}\indic{\pi_{i^*}>0}\indic{a_j>0,\forall j\in L_{i^*}}\right)\right)Q_{i^*}(n)\indic{Q_{i^*}(n)>0}\nonumber\\
		&\leq
	-m\, \left(\min_i\max_{\mf{a}\in\{0,1\}^K}\max_{\bm{\pi}\in\mc{M}}\mb{P}(\mf{T}(n)=\mf{a})\left( q_{i}\indic{\pi_{i}>0}\indic{a_j>0,\forall j\in L_{i}}\right)\right)Q_{i^*}(n)\indic{Q_{i^*}(n)>0}\nonumber\\
	&\leq
	-mm'\frac{\norm{\mf{Q}(n)}}{{\sqrt{M}}},
	\end{align}
	where $m'=\min_i\max_{\mf{a}\in\{0,1\}^K}\max_{\bm{\pi}\in\mc{M}}\mb{P}(\mf{T}(n)=\mf{a})\left( q_{i}\indic{\pi_{i}>0}\indic{a_j>0,\forall j\in L_{i}}\right)$ and we assume $Q_{i^*}(n)>0$ in the last equation. Finally, we conclude the result by choosing $\epsilon=\frac{mm'}{\sqrt{M}}$.
 \end{proof}
	
	We establish the stability of $(\mf{Q}(n),n\geq 0)$ using a Lyapunov stability theorem \citep{Tweedle}. We use the real valued Lyapunov function $V((\mf{Q}(n)))=\sum_{i=1}^MQ_i^2(n)$ to show
	\beq
	\mb{E}(V(\mf{Q}(n+1))-V(\mf{Q}(n))\given \mf{Q}(n))<-\epsilon' \norm{\mf{Q}(n)},
	\eeq
	 for sufficiently large $\norm{\mf{Q}(n)}$, where $\epsilon'>0$.
	\begin{theorem}
	If $\bm{\lambda}\in\Lambda_Q$ and $\mb{E}[\sum_{i=1}^MA^2_i(n)]<\infty$, the Max-Weight scheduling policy stabilizes the quantum network.	
\end{theorem}
\begin{proof}	
	We can write 
	\beq
	V(\mf{Q}(n))=\mf{Q}(n)\cdot \mf{Q}(n).
	\eeq
	Furthermore, we can write
	\begin{align}
	   V(\mf{Q}(n+1))-V(\mf{Q}(n))&=\mf{Q}(n+1)\cdot \mf{Q}(n+1)-\mf{Q}(n)\cdot\mf{Q}(n)\nonumber\\
	   &=(\mf{Q}(n+1)-\mf{Q}(n))\cdot (\mf{Q}(n+1)-\mf{Q}(n))+2(\mf{Q}(n+1)-\mf{Q}(n))\cdot\mf{Q}(n)
	   \end{align}
	   Hence, we obtain
	   \begin{multline}
	   \expect{V(\mf{Q}(n+1))-V(\mf{Q}(n))\given \mf{Q}(n),\mf{T}(n)}=\expect{(\mf{Q}(n+1)-\mf{Q}(n))\cdot (\mf{Q}(n+1)-\mf{Q}(n))\given \mf{Q}(n),\mf{T}(n)}\\
	   +2\expect{(\mf{Q}(n+1)-\mf{Q}(n))\cdot\mf{Q}(n)\given \mf{Q}(n),\mf{T}(n)}\label{eq:deviation}
	   \end{multline}
	   The first term in the right side of the above equation can be bounded by
	   \begin{align}
	   \expect{(\mf{Q}(n+1)-\mf{Q}(n))\cdot (\mf{Q}(n+1)-\mf{Q}(n))\given \mf{Q}(n),\mf{T}(n)}&\leq \expect{\sum_{i=1}^M(A_i(n)+1)^2}\nonumber\\
	   &=\expect{\sum_{i=1}^MA_i^2(n)}+M+2\sum_{i=1}^M\lambda_i.\label{eq:deviation2}
	   \end{align}
	   
	   Next we simplify the second term in the right side of \eqref{eq:deviation}. We can write
	   \beq
	   \expect{(\mf{Q}(n+1)-\mf{Q}(n))\cdot\mf{Q}(n)\given \mf{Q}(n),\mf{T}(n)}\nonumber
	   \eeq\begin{align}
	   &=\expect{(\mf{A}(n)\cdot\mf{Q}(n)\given \mf{Q}(n),\mf{T}(n)}-\expect{(\mf{D}(n)\cdot\mf{Q}(n)\given \mf{Q}(n),\mf{T}(n)}\nonumber\\
	   &=\bm{\lambda}\cdot\mf{Q}(n)-\max_{\bm{\pi}\in\mc{M}}\mf{u}(\mf{Q}(n),\mf{T}(n),\bm{\pi})\cdot\mf{Q}(n).\label{eq:deviation3}
	   \end{align}
	   We now obtain the following equation by substituting equations  \eqref{eq:deviation2} and \eqref{eq:deviation3} in \eqref{eq:deviation}
	   \begin{multline}
	   \expect{V(\mf{Q}(n+1))-V(\mf{Q}(n))\given \mf{Q}(n),\mf{T}(n)}\leq\expect{\sum_{i=1}^MA_i^2(n)}+M+2\sum_{i=1}^M\lambda_i\\
	   +2(\bm{\lambda}\cdot\mf{Q}(n)-\max_{\bm{\pi}\in\mc{M}}\mf{u}(\mf{Q}(n),\mf{T}(n),\bm{\pi})\cdot\mf{Q}(n)).
	   \end{multline}
	   Next we take expectation over $\mf{T}(n)$ in the above equation to obtain
	   \begin{multline}
	   \expect{V(\mf{Q}(n+1))-V(\mf{Q}(n))\given \mf{Q}(n)}\leq\expect{\sum_{i=1}^MA_i^2(n)}+M+2\sum_{i=1}^M\lambda_i\\
	   +2(\bm{\lambda}\cdot\mf{Q}(n)-\sum_{\mf{a}\in\{0,1\}^K}\mb{P}(\mf{T}(n)=\mf{a})\max_{\bm{\pi}\in\mc{M}}\mf{u}(\mf{Q}(n),\mf{a},\bm{\pi})\cdot\mf{Q}(n)).
	   \end{multline}
	   Using Lemma~\ref{thm:gap_bound_lemma}, we get
	   \beq
	   \expect{V(\mf{Q}(n+1))-V(\mf{Q}(n))\given \mf{Q}(n)}\leq\expect{\sum_{i=1}^MA_i^2(n)}+M+2\sum_{i=1}^M\lambda_i
	   -2\epsilon\norm{\mf{Q}(n)}.
	   \eeq
	   Finally, if $\norm{\mf{Q}(n)}\geq \frac{2(\expect{\sum_{i=1}^MA_i^2(n)}+M+2\sum_{i=1}^M\lambda_i)}{\epsilon}$, we get
	  \beq
	   \expect{V(\mf{Q}(n+1))-V(\mf{Q}(n))\given \mf{Q}(n)}\leq
	   -\frac{\epsilon}{2}\norm{\mf{Q}(n)}.
	   \eeq 
	   This completes the proof.
	   
\end{proof}

	\section{Numerical Results}
	\label{sec:numerics}
	In this section, we provide numerical results on the analysis of quantum networks. We study the impact of the Max-Weight scheduling policy on queue sizes of requests. For numerical results, we assume that one type~$i$ request arrives with probability $\lambda_i$ and no type~$i$ request arrives with probability $1-\lambda_i$ in each time-slot.
	
	First, we consider a single switch with four links ($K=4$) that aims to create bipartite entanglements for user-pairs. There are six types of requests arriving in the system serving user-pairs $U_1=\{1,2\}$, $U_2=\{1,3\}$, $U_3=\{1,4\}$, $U_4=\{2,3\}$, $U_5=\{2,4\}$, and $U_6=\{3,4\}$. The arrival rates of requests of different types are: $\lambda_1=0.3$, $\lambda_2=0.3$, $\lambda_3=0.3$, $\lambda_4=0.3$, $\lambda_5=0.45$, and $\lambda_6=0.2$. It can be verified that $\bm{\lambda}\notin\Lambda_Q$ and the corresponding partition of $\bm{\lambda}$ is given by
	\begin{multline}
			\bm{\lambda}=0.2\begin{bmatrix}
				1 & 0 &  0 &  0&0&1
			\end{bmatrix}+0.2\begin{bmatrix}
			0 & 0 &  1 &  1&0&0
		\end{bmatrix}+0.3\begin{bmatrix}
		0 & 1 &  0 &  0&1&0
	\end{bmatrix}\\
			+0.1\begin{bmatrix}
				1 & 0 &  0 &  0&0&0
			\end{bmatrix}
			+0.1\begin{bmatrix}
				0 & 0 &  1 &  0&0&0
			\end{bmatrix}+0.15\begin{bmatrix}
			0 & 0 &  0 &  0&1&0
		\end{bmatrix}.
		\end{multline}
	Hence, the switch should be unstable under any scheduling policy.
	
\begin{figure}
		\centering
		\includegraphics[scale=1,width=0.65\textwidth]{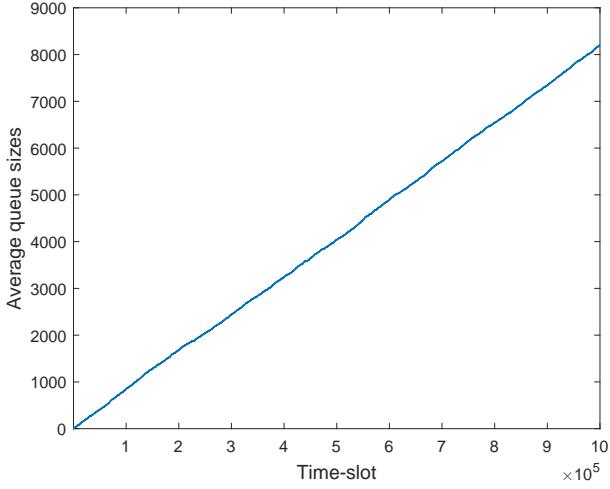}
		\caption{Time evolution of average queue sizes of requests in an unstable switch}
		\label{fig:unstable_queue}
	\end{figure}

\begin{figure}
		\centering
		\includegraphics[scale=1,width=0.65\textwidth]{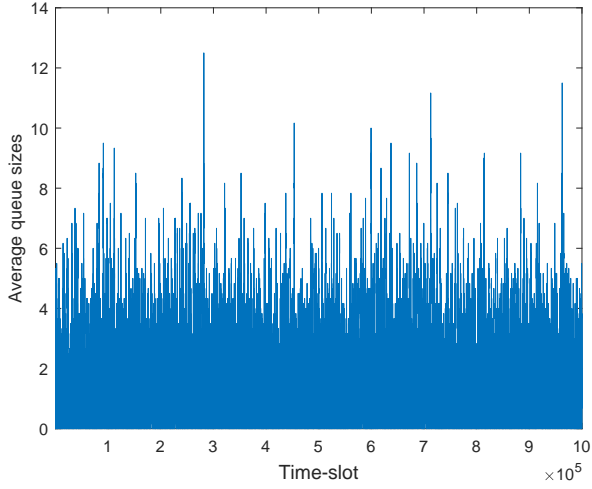}
		\caption{Time evolution of average queue sizes of requests in a stable switch}
		\label{fig:stable_queue}
	\end{figure}

	In Figures~\ref{fig:unstable_queue} and \ref{fig:stable_queue}, we simulate the switch under the assumption that it uses our Max-Weight scheduling policy. In Figure~\ref{fig:unstable_queue}, we plot the time evolution of average queue sizes of requests denoted by $S(n)$ as a function of $n$,
	where
	\beq
	S(n)=\frac{\sum_{i=1}^MQ_i(n)}{M}.
	\eeq
	It is clear that the average queue sizes of requests increase monotonically with $n$, imply that the switch is unstable.

	Next, we simulate a stable switch with the assumption that the set of arrival rates are: $\lambda_1=0.3$, $\lambda_2=0.3$, $\lambda_3=0.2$, $\lambda_4=0.3$, $\lambda_5=0.45$, and $\lambda_6=0.2$. It can be verified that $\bm{\lambda}\in\Lambda_Q$ and the corresponding partition of $\bm{\lambda}$ is given by
	\begin{multline}
			\bm{\lambda}=0.2\begin{bmatrix}
				1 & 0 &  0 &  0&0&1
			\end{bmatrix}+0.2\begin{bmatrix}
			0 & 0 &  1 &  1&0&0
		\end{bmatrix}+0.3\begin{bmatrix}
		0 & 1 &  0 &  0&1&0
	\end{bmatrix}\\
			+0.1\begin{bmatrix}
				1 & 0 &  0 &  0&0&0
			\end{bmatrix}
			+0.15\begin{bmatrix}
			0 & 0 &  0 &  0&1&0
		\end{bmatrix}.
		\end{multline}
Hence, the switch should be stable under our Max-Weight scheduling policy.
In Figure~\ref{fig:stable_queue}, we plot the average queue sizes of requests $S(n)$ as a function of $n$. We observe that the average queue sizes of requests are finite even at large values of $n$. Hence, the switch is stable.	

	Now we simulate the quantum network shown in Figure~\ref{fig:switch_network}. Suppose that there are two types of requests with $U_1=\{1,2,3\}$, $U_2=\{3,4,5\}$, $L_1=\{l_1,l_2,l_3,l_6\}$, and $L_2=\{l_3,l_4,l_5,l_7\}$. Note that the link $l_3$ is used by both types of requests, hence, only one type of request can be served in each time-slot. We will compare the performance of the Max-Weight scheduling policy with that of the policy referred to as Policy 2 that chooses the type of the request in each time-slot as the type with maximum queue sizes and ties are broken uniformly at random.

\begin{figure}
		\centering
		\includegraphics[scale=1,width=0.65\textwidth]{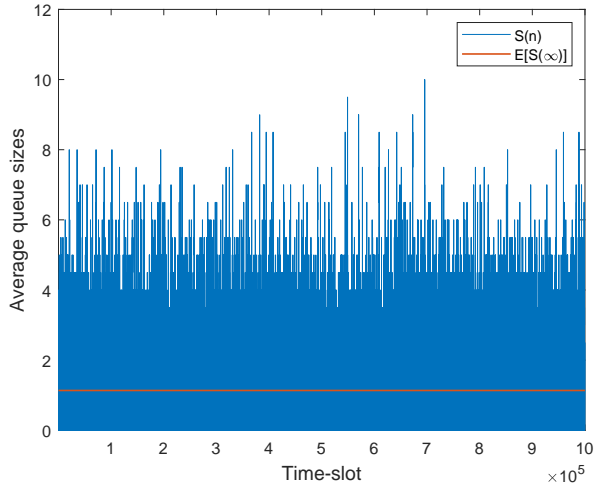}
		\caption{Time evolution of average queue sizes of requests under the Max-Weight scheduling with $\bm{\lambda}=[0.095 \,\,0.165]$}
		\label{fig:stable_network1}
	\end{figure}

\begin{figure}
		\centering
		\includegraphics[scale=1,width=0.65\textwidth]{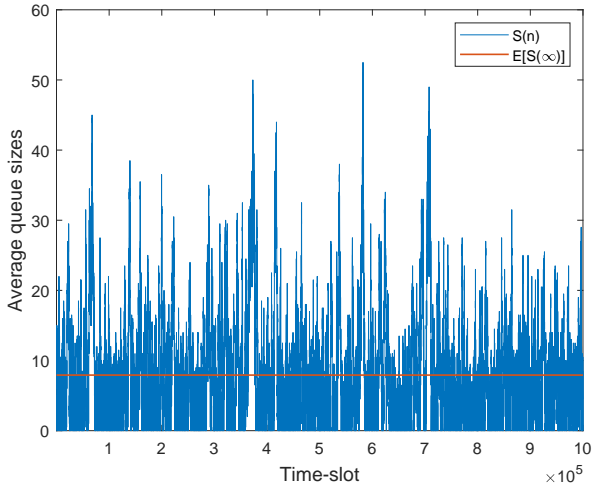}
		\caption{Time evolution of average queue sizes of requests under Policy 2 with $\bm{\lambda}=[0.095 \,\,0.165]$}
		\label{fig:stable_network_max}
	\end{figure}

	For Figures~\ref{fig:stable_network1} and \ref{fig:stable_network_max}, we choose the arrival request rates as $\bm{\lambda}=[0.095 \,\,0.165]$, the link-level entanglement generation probabilities are $\mf{p}=[0.7\,\, 0.8\,\, 0.6\,\, 0.9 \,\,0.9\,\, 0.9\,\, 0.8]$, and the measurement success probabilities are $\mf{q}=[0.75 \,\,0.8]$. We plot the time evolution of average queue sizes of requests $S(n)$ as a function of $n$. We also plot $\mb{E}[S(\infty)]$ as a function of $n$, where
	\beq
	\mb{E}[S(\infty)]=\lim_{t\to\infty}\frac{1}{t}\sum_{n=1}^tS(n).
	\eeq
	In our numerical results, we choose $t=10^6$.
	The Figure~\ref{fig:stable_network1} corresponds to the Max-Weight scheduling policy and the Figure~\ref{fig:stable_network_max} corresponds to Policy 2. It is clear that the stationary average queue size  $\mb{E}[S(\infty)]$ under the Max-Weight policy is smaller than the corresponding stationary average queue size $\mb{E}[S(\infty)]$ under Policy~2.

\begin{figure}
		\centering
		\includegraphics[scale=1,width=0.65\textwidth]{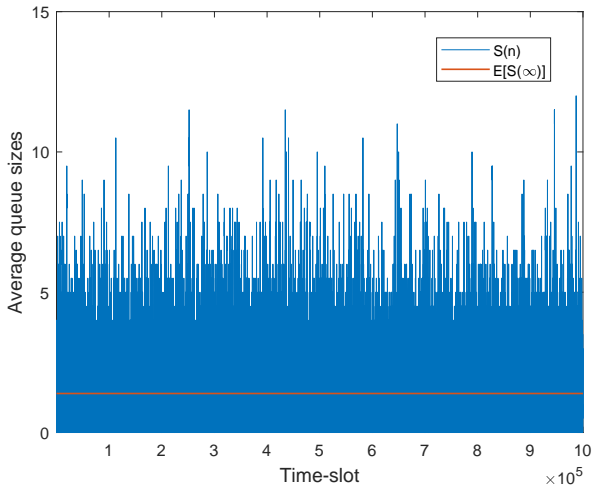}
		\caption{Time evolution of average queue sizes of requests under the Max-Weight scheduling with $\bm{\lambda}=[0.105 \,\,0.175]$}
		\label{fig:stable_network2}
	\end{figure}

\begin{figure}
		\centering
		\includegraphics[scale=1,width=0.65\textwidth]{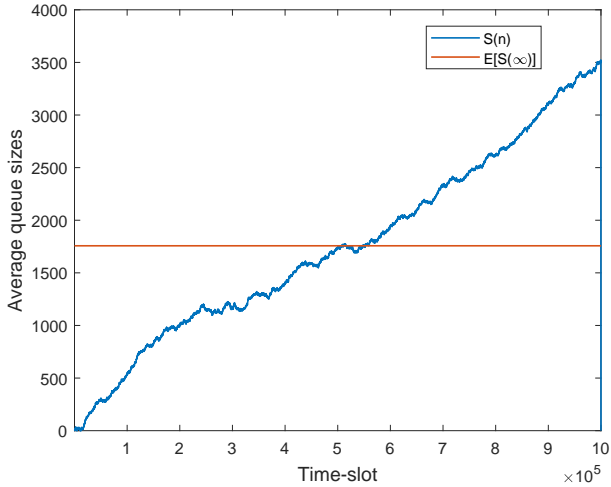}
		\caption{Time evolution of average queue sizes of requests under Policy 2 with $\bm{\lambda}=[0.105 \,\,0.175]$}
		\label{fig:unstable_network_max}
	\end{figure}

	We now repeat the experiments by choosing the same parameters except now we choose $\bm{\lambda}=[0.105 \,\,0.175]$. The Figure~\ref{fig:stable_network2} corresponds to the Max-Weight scheduling policy and the Figure~\ref{fig:unstable_network_max} corresponds to Policy 2. It is clear that the network is stable under the Max-Weight scheduling policy and it is unstable under Policy 2 as the average queue sizes increase monotonically with $n$. This example shows that we should design a scheduling policy by considering dynamic properties of quantum networks into account to have good efficiency.
	

	\section{Conclusions}
\label{sec:conclusions}
In this paper, we studied entanglement distribution on quantum networks that aim to distribute entanglements for different sets of users. First, we investigated stability properties of quantum networks by deriving necessary conditions on request arrival rates for achieving stability of the network. By exploiting the fact that the generation of link-level entanglements and measurement operations are successful with certain probabilities, we proposed a Max-Weight scheduling policy and were able to show that the proposed policy stabilizes the network for all feasible arrival rates.

A major drawback of our policy is that it has large implementation cost due to the fact that we need to search over all possible matchings to find the best matching in each time-slot.
In the future, it is of interest to find other low complex scheduling policies that can stabilize the network for all feasible arrival rates. We would also like to investigate the model in which links can create multiple entanglements in each time-slot. This model will have larger capacity regions compared to the model addressed in this paper. Finally, in this paper, we assumed that each type of requests can be served by using a single set of links. We plan to extend our analysis to the setting where each type of requests can be served by several sets of links.
	
\begin{acks}
This work was supported by the National Science Foundation (NSF) under Grants  CNS-195744 and  ERC-1941583. 
\end{acks}

\bibliographystyle{ACM-Reference-Format}
\bibliography{reference_performance21}


	\appendix
	\label{sec:appendix}
	\section{Proof of Theorem~\ref{thm:stable}}
	\label{app:stable}
	The proof uses the existence of the stationary distribution under a scheduling policy. First, we begin with the simple case of $p_i=1$ for $1\leq i\leq K$ and $q_j=1$ for $1\leq j\leq M$.
	\subsection{For $p_i=1$ $(1\leq  i\leq K)$ and $q_j=1$ ($1\leq j\leq M$) :}
	In this section, we assume that every link generates an entanglement in each time-slot and the entanglement measurement operations are always successful. As a consequence,
	the number of served requests of type~$i$ in time-slot $n$ is given by
	\beq
	D_{i}(n)=\indic{Q_{i}(n)>0}\indic{W_{i}(n)>0}.
	\eeq
	Furthermore, we have
	\beq
	\label{eq:que_dyn2}
	\mf{Q}(n+1)=\mf{Q}(1)-\sum_{j=1}^n\mf{D}(j)+\sum_{j=1}^n\mf{A}(j).
	\eeq
	We can also write
	\beq
	\sum_{j=1}^n\mf{D}(j)=\sum_{\bm{\sigma}\in\mc{D}}c_{\bm{\sigma}}(n)\bm{\sigma}\text{ and }\sum_{\bm{\sigma}\in\mc{D}}c_{\bm{\sigma}}(n)=n,
	\eeq
	where $c_{\bm{\sigma}}(n)$ denotes the cumulative number of times the vector $\bm{\sigma}$ is served up to time-slot $n$. 
	Hence,
	\begin{align}
		\frac{\mf{Q}(n+1)}{n}&=\frac{\mf{Q}(1)}{n}-\frac{\sum_{j=1}^n\mf{D}(j)}{n}+\frac{\sum_{j=1}^n\mf{A}(j)}{n}\nonumber\\
		&=\frac{\mf{Q}(1)}{n}-\sum_{\bm{\sigma}\in\mc{D}}\frac{c_{\bm{\sigma}}(n)}{n}\bm{\sigma}+\frac{\sum_{j=1}^n\mf{A}(j)}{n}.
	\end{align}
	Since $\expect{\mf{Q}(1)}<\infty$, we obtain
	\beq
	\label{eq:q1}
	\lim_{n\to\infty}\frac{\mf{Q}(1)}{n}=0.
	\eeq
	The elements of the sequence $\{A_{i}(n)\}$  are independent and identically distributed (iid) random variables, hence, we have
	\beq
	\label{eq:arr_rate}
	\lim_{n\to\infty}\frac{\sum_{j=1}^n\mf{A}(j)}{n}
	=\bm{\lambda}.
	\eeq
	Since the system is assumed to be stable, we have
	\beq
	\label{eq:qn_stable}
	\lim_{n\to\infty}\frac{\mf{Q}(n+1)}{n}=0.
	\eeq
	Therefore, we get
	\begin{align}
		\lim_{n\to\infty}\frac{\sum_{j=1}^n\mf{D}(j)}{n}=\bm{\lambda}.
	\end{align}
	The above equation shows that the rate of departure of requests at equilibrium coincides with the arrival rate of requests $\bm{\lambda}$.
	Hence, we obtain
	\beq
	\sum_{\bm{\sigma}\in\mc{D}}\lim_{n\to\infty}\frac{c_{\bm{\sigma}}(n)}{n}\bm{\sigma}=\bm{\lambda}.
	\eeq
	
	It remains to show that the limit $\lim_{n\to\infty}\frac{c_{\bm{\sigma}}(n)}{n}$ exists. By definition, we have
	\beq
	c_{\bm{\sigma}}(n)=\sum_{j=1}^n\indic{\mf{D}(j)=\bm{\sigma}}.
	\eeq
	Consider the Markov chain $(\mf{Q}(n),\mf{Y}(n))$ where $\mf{Y}(n)$ is a random variable that is used to break ties if multiple matchings can be chosen according to the Max-Weight scheduling policy. Let $\mc{C}_{\bm{\sigma}}$ be the set of states of $(\mf{Q}(n),\mf{Y}(n))$ for which the switch serves the matching $\bm{\sigma}$ in time-slot $n$. Then we write
	\beq
	c_{\bm{\sigma}}(n)=\sum_{(\mf{q},y)\in \mc{C}_{\bm{\sigma}}}\sum_{j=1}^n\indic{\mf{Q}(j)=\mf{q},\mf{Y}(j) =y}.
	\eeq
	Then it follows that the following limit exists due to the stability of the system
	\beq
	\lim_{n\to\infty}\frac{c_{\bm{\sigma}}(n)}{n}=c^*_{\bm{\sigma}},
	\eeq
	where $c^*_{\bm{\sigma}}$ denotes the probability with which the matching $\bm{\sigma}$ is served in a time-slot at equilibrium. Since we have
	\beq
	\sum_{\bm{\sigma}\in\mc{D}}\frac{c_{\bm{\sigma}}(n)}{n}=1.
	\eeq
	We conclude that $\sum_{\bm{\sigma}\in\mc{D}}c^*_{\bm{\sigma}}=1$.
	
	\subsection{For $p_i\leq 1$ $(1\leq  i\leq K)$ and $q_j\leq 1$ ($1\leq j\leq M$) :}
	
	A matching $\mf{W}(n)$ is selected in time-slot $n$ according to the Max-Weight scheduling as in \eqref{eq:max_weight}.
	The number of type~$i$ requests served in time-slot $n$ is given by
	\beq
	D_{i}(n)=Z_{i}(n)\indic{Q_{i}(n)>0}\indic{T_{j}(n)>0,\forall j\in L_i}\indic{W_{i}(n)>0}.
	\eeq
	Since $\expect{\mf{Q}(1)}<\infty$ and the system is stable, the equations \eqref{eq:q1}, \eqref{eq:arr_rate}, and \eqref{eq:qn_stable} are valid.
	As a result, we get
	\begin{align}
		\lim_{n\to\infty}\frac{\sum_{j=1}^n\mf{D}(j)}{n}=\bm{\lambda}.
	\end{align}
	The above equation shows that the rate of departure of requests at equilibrium coincides with the arrival rate of requests $\bm{\lambda}$.
	Hence, we obtain
	\beq
	\sum_{\bm{\sigma}\in\mc{D}}\lim_{n\to\infty}\frac{c_{\bm{\sigma}}(n)}{n}\bm{\sigma}=\bm{\lambda}.
	\eeq
	
	It remains to show that the limit $\lim_{n\to\infty}\frac{c_{\bm{\sigma}}(n)}{n}$ exists. By definition, we have
	\beq
	c_{\bm{\sigma}}(n)=\sum_{j=1}^n\indic{\mf{D}(j)=\bm{\sigma}}.
	\eeq
		Consider the Markov chain $(\mf{Q}(n),\mf{T}(n),\mf{Y}(n),\mf{Z}(n))$ where $\mf{Z}(n)=(Z_{i}(n))$, where the random variable $\mf{Y}(n)$ is used to break ties if multiple matchings can be chosen according to the Max-Weight scheduling policy for given $\mf{Q}(n)$ and $\mf{T}(n)$. Let $\mc{C}_{\bm{\sigma}}$ be the set of states of $(\mf{Q}(n),\mf{T}(n),\mf{Y}(n),\mf{Z}(n))$ for which the systems serves the matching $\bm{\sigma}$ in time-slot $n$. Then we write
	\beq
	c_{\bm{\sigma}}(n)=\sum_{(\mf{q},\mf{t},y,\mf{z})\in \mc{C}_{\bm{\sigma}}}\sum_{j=1}^n\indic{\mf{Q}(j)=\mf{q},\mf{T}(j)=\mf{t},\mf{Y}(j) =y, \mf{Z}(j)=\mf{z}}.
	\eeq
	Then it follows that since the system is stable
	\beq
	\lim_{n\to\infty}\frac{c_{\bm{\sigma}}(n)}{n}=c^*_{\bm{\sigma}},
	\eeq
	where $c^*_{\bm{\sigma}}$ denotes the probability with which the vector $\bm{\sigma}$ is served in a time-slot at equilibrium. Further, we have
	\beq
	\label{eq:arr_partition}
	\bm{\lambda}=\sum_{\bm{\sigma}\in\mc{D}}c^*_{\bm{\sigma}}\bm{\sigma}.
	\eeq
	
	
	From \eqref{eq:arr_partition}, we have
	\begin{align}
		\lambda_{i}&=\sum_{\bm{\sigma}\in\mc{D}}c^*_{\bm{\sigma}}\sigma_{i}\nonumber\\
		&=\sum_{\bm{\sigma}\in\mc{D},\sigma_{i}=1}c^*_{\bm{\sigma}}.
	\end{align}
	Hence, we obtain
	\begin{align}
		\lambda^*_{i}=\frac{\lambda_{i}}{q_i\prod_{j\in L_i}p_j}
		&=\sum_{\bm{\sigma}\in\mc{D}}\frac{c^*_{\bm{\sigma}}}{q_i\prod_{j\in L_i}p_j}\sigma_{i}\nonumber\\
		&=\sum_{\bm{\sigma}\in\mc{D}}r^*_{i}(\bm{\sigma})\sigma_{i},\label{eq:nec_derived}
	\end{align}
	where $r^*_{i}(\bm{\sigma})=\frac{c^*_{\bm{\sigma}}}{q_i\prod_{j\in L_i}p_j}$. We can interpret $r^*_{ij}(\bm{\sigma})$ as the stationary probability with which $\bm{\sigma}$ is served in a time-slot conditioned on the event that all the links $L_i$ have qubits and all the  entanglement measurement operations are successful.\qed

\end{document}